# Time development of a driven three-level lambda system: A case study

**James P. Lavine**, Department of Physics, Georgetown University, 37th and O Sts. NW, Washington, D.C. 20057, United States of America

E-mail: jpl62@georgetown.edu and lavine.james@gmail.com

**Abstract**

How does a driven system with many energy levels approach its steady state? Insights are gained by studying a system with three energy levels when the ground state is excited by a laser. The time-dependent occupation probabilities of the three energy levels show students how the system develops in time. The occupation probabilities come from the numerical solution of the Liouville-von Neumann equations for the density operator matrix elements when relaxation is included. A combination of the Interaction Picture, the Rotating Wave Approximation, and the assumption of resonance permit the eigenvalues of the Liouville-von Neumann equations to be found numerically and in closed-form in certain limits. The two methods are complementary and help students understand time-dependent systems. In addition, the eigenvalues allow the short-time and the long-time occupation probabilities to be connected to the relaxation parameters and the magnitude of the laser's electric field. Thus, this model three-level system illuminates how a driven system behaves over time and provides guidance for students studying time-dependent systems.

## 1. Introduction

A dynamical system is described by its initial state and its time-dependent governing equations. When the time derivatives in the governing equations are set to zero, the steady-state values of the system may be obtained. How the system evolves from its initial values to its steady state generally requires numerical and/or experimental work. This is a challenging problem in non-equilibrium Physics [1]. Solutions for specific systems help to develop insights that may be applicable to more general systems.

Here a particular three-level system is assumed with a ground state and two excited states. The three energy levels form an inverted V or lambda, Λ, shape as shown in figure 1. A laser drives transitions between the ground state and the highest excited state. Relaxation mechanisms between the energy levels are indicated in figure 1. In such a system, an electron in the highest energy level, energy level 1, may decay to energy level 0, the ground state, or to energy level 2, another excited state, which in turn decays to the ground state. The present investigation seeks the time-dependent occupation probabilities of the three energy levels. The key questions addressed are: what is the short-time behavior, where the majority of the occupation probability resides in the steady state, and how the steady state is reached in time.

The time development of the occupation probabilities of the three energy levels is followed from the time $t = 0$ to the steady state. This is done through the numerical solution of the Liouville-von Neumann equations for the matrix elements of the density operator. These equations are derived from the definition of the density operator and Schrödinger's equation [2]. This approach allows simple relaxation mechanisms to be taken into account and is supplemented by consideration of the eigenvalues of the Liouville-von Neumann equations.

This combination of methods probes the time constants that govern the occupation probabilities at short times and their entrance to steady state. In fact, the role of the relaxation parameters and the magnitude of the laser's electric field are exposed. Thus, this model three-level system serves as an illustration of how a multi-level system behaves over time.

The Liouville-von Neumann equations are formulated in the Interaction Picture [2, 3] and the Rotating Wave Approximation [4] is made. When the laser's energy is equal to the energy difference between energy level 1 and the ground state, resonance is achieved. The coefficients in the right-hand side of the Liouville-von Neumann equations are now independent of time as shown in Appendix A. This has two consequences. Firstly, the steady-state values of the occupation probability are obtained by algebra. The results show non-zero values for the occupation probabilities of the ground state, energy level 0, and both of the excited states. Secondly, the eigenvalues of the Liouville-von Neumann equations are independent of time. These eigenvalues go into time-dependent exponentials whose sum is an occupation probability. Each occupation probability involves the same exponentials. Hence, non-zero steady-state values require a zero eigenvalue and this is explored in Appendix B.

The three-level lambda or inverted V system is often used as an approximation to a multi-level physical system and is discussed by Berman and Malinovsky [4], Shore [5], and Scully and Zubairy [6]. Sanchez and Brandes [7] include dissipation and study the loss of coherent population trapping. Manka, Doss, Narducci, Ru, and Oppo [8] and Blaauboer [9] study lambda systems driven by two lasers and touch on the steady state. Rose, Popolitova, Tikhonova, Meir, and Sharapova [10] and Sen, Dey, Nath, and Gangopadhyay [11] are recent works that focus on dark states [4, 6] in the lambda system.

The present study uses one laser that excites only one transition. The concern is with how the occupation probabilities change with time and how the results relate to the system parameters for this case. The second section presents the Liouville-von Neumann equations for the density operator matrix elements, $\rho_{ij}(t)$, with relaxation. Appendix A has the derivation of the equations used here for the $\rho_{ij}$ when relaxation is ignored. The third section has the steady-state solution and introduces the eigenvalues of the Liouville-von Neumann equations for the present case. The eigenvalues themselves are developed in Appendix B. The fourth section has the numerical solutions of the time-dependent, coupled Liouville-von Neumann equations of the second section. The results include how the density operator matrix elements, especially the occupation probabilities, behave at short times and how they approach the steady state. The eigenvalues are then applied to the short-time and the long-time behavior in order to relate the slopes of the occupation probabilities to the parameters of the model. The final section has the conclusions.

## 2. The Liouville-von Neumann Equations

The full Liouville-von Neumann Equations for a three-level system are a coupled set of nine first-order, linear, ordinary differential equations for the nine matrix elements of the density operator $\rho$. Relaxation is first ignored and the Liouville-von Neumann equation in the Interaction Picture in operator form is [12]

$$i\hbar\, d\rho(t)/dt = [V(t), \rho(t)] \ . \tag{1}$$

This equation is developed into the nine equations for the density matrix elements in Appendix A, which demonstrates that four of the matrix elements decouple for the lambda system of figure

1. The remaining five matrix elements are coupled and they include the occupation probabilities $\rho_{ii}(t)$ for i = 0 to 2. The Rotating Wave Approximation [2] is also explained in Appendix A. Figure 1 shows the allowed transitions included here when relaxation is added. Formal derivations of this approach are found in Blum [13]. A laser is used to excite the ground state, energy level 0, to the excited state, energy level 1, while relaxation allows the transfers from energy levels 1 and 2 to the ground state, energy level 0. The former has $k_{01} = 1/T_1$ and $k_{21}$ for energy level 1 to the ground state and to energy level 2, respectively. The rate for energy level 2 to the ground state is $k_{02}$. Finally, the off-diagonal matrix elements approach the steady-state with a relaxation time $T_2$.

The laser energy equals the energy difference between energy levels 1 and 0, so resonance is assumed. As shown in Appendix A, resonance and the Rotating Wave Approximation result in interaction matrix elements that are independent of time. All of this leads to four coupled equations that resemble those of Basché et al. [14] for the matrix elements of the density matrix $\rho_{ij}$ as functions of time,

$$d\rho_{00}(t)/dt = -\big(\Omega\rho_B(t)\big) + (1/T_1)\rho_{11}(t) + k_{02}\rho_{22}(t)\ , \tag{2}$$

$$d\rho_B(t)/dt = (\Omega/2)\rho_{00}(t) - (1/T_2)\rho_B(t) - (\Omega/2)\rho_{11}(t)\ , \tag{3}$$

$$d\rho_{11}(t)/dt = \big(\Omega\rho_B(t)\big) - (1/T_1)\rho_{11}(t) - k_{21}\rho_{11}(t)\ \ , \tag{4}$$

$$d\rho_{22}(t)/dt = k_{21}\,\rho_{11}(t) - k_{02}\rho_{22}(t)\ \ . \tag{5}$$

These relaxation terms lead to exponential decay in time when the other terms on the right-hand sides are ignored. The $\rho_B$ comes from the imaginary parts of the off-diagonal matrix elements, $\rho_B = (1/i)(\rho_{01} - \rho_{10})$, while the real parts decouple. Equation (5) shows the occupation probability of energy level 2 changes due to relaxation in accord with figure 1. In addition,

$$\Omega = E\mu/\hbar\ , \tag{6}$$

with $E$ the magnitude of the laser's electric field, $\mu$ the electric dipole moment between energy levels 0 and 1, and $\hbar$ Planck's constant divided by $2\pi$. The angular frequencies are taken to be in units of $10^9$ radians/s and the time in nanoseconds. The sum of equations (2), (4), and (5) says the total probability is conserved, so

$$\rho_{00}(t) + \rho_{11}(t) + \rho_{22}(t) = 1.0\ , \tag{7}$$

and this serves as a check on the numerical solutions.

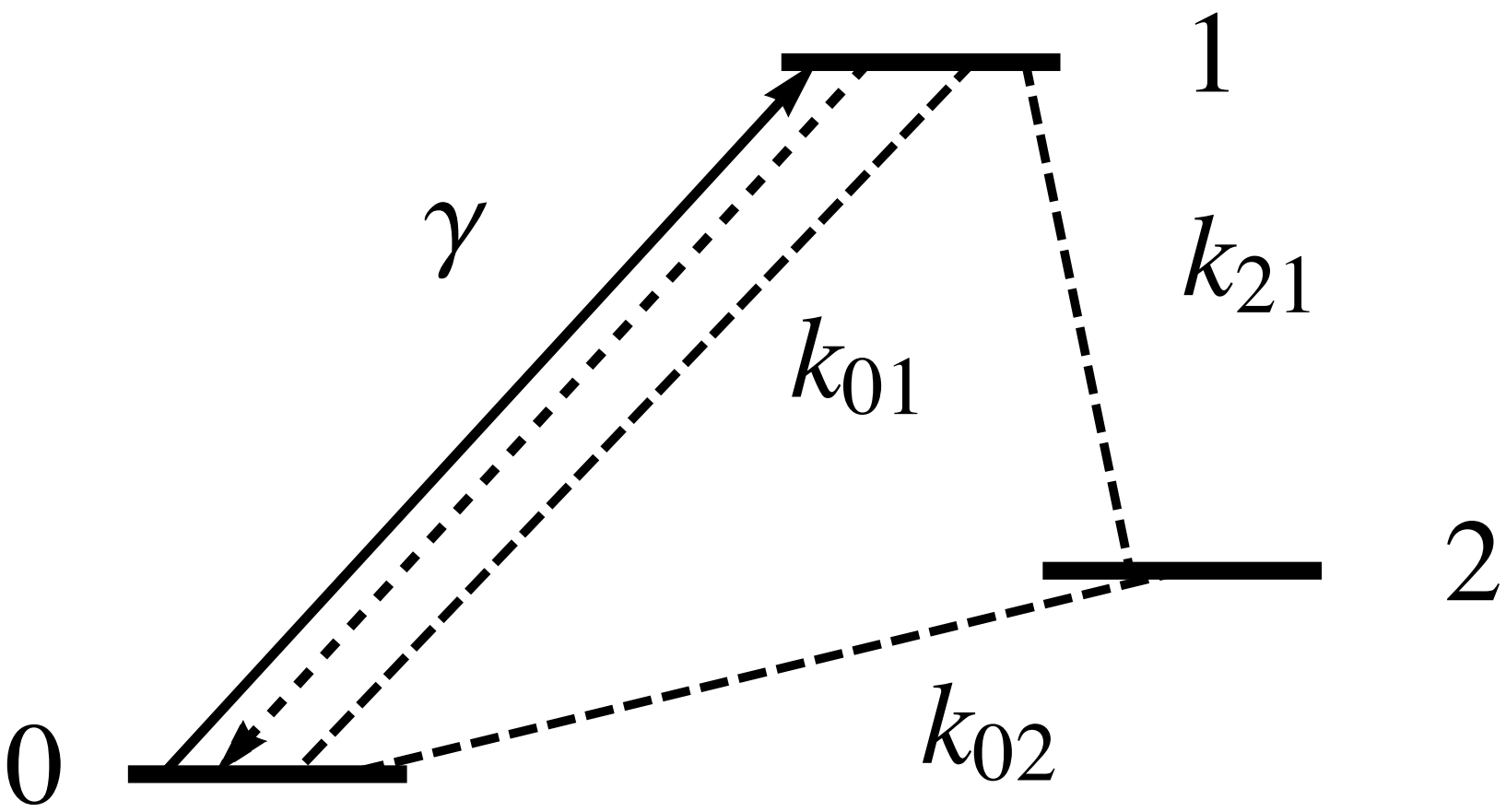

**Figure 1.** Schematic of a three-level lambda system. Photons, $\gamma$, excite the system from the ground state to energy level 1. The $k_{ij}$ represent relaxation paths and the dashed line with the arrowhead is stimulated emission. Please note $k_{01} = {}^{1}/_{T_1}$.

The next section finds the steady-state values for the $\rho_{ij}$ of equations (2) to (5).

## 3. Steady-State Solution and the Eigenvalues

The steady-state values are found by setting the time derivatives of equations (2) to (5) to zero. Equation (5) says

$$\rho_{22}(\infty) = \frac{k_{21}}{k_{02}} \rho_{11}(\infty) \quad . \tag{8}$$

Then equation (4) yields

$$\rho_B(\infty) = \frac{1}{\Omega}\left(\frac{1}{T_1} + k_{21}\right) \rho_{11}(\infty) \quad . \tag{9}$$

Now equation (3) leads to

$$\rho_{00}(\infty) = \rho_{11}(\infty) + \frac{2}{\Omega T_2} \rho_B(\infty) \quad , \tag{10}$$

and with equation (9), $\rho_{00}(\infty)$ is in terms of only $\rho_{11}(\infty)$. Next,

$$\rho_{00}(\infty) + \rho_{11}(\infty) + \rho_{22}(\infty) = 1 \ , \tag{11}$$

along with equations (8) and (9) give

$$\rho_{11}(\infty) = (\Omega^2/2) / \left[\Omega^2 \left(1 + \frac{k_{21}}{2k_{02}}\right) + \frac{1}{T_1 T_2} + \frac{k_{21}}{T_2}\right] \quad . \tag{12}$$

All the $\rho_{ii}(\infty)$ and $\rho_B(\infty)$ are now available as functions of the parameters of the Liouville-von Neumann equations.

The present study uses

$$T_1 = 0.277/3 \ = 0.0923333 \ ns \ ,$$

$$T_2 = 0.132 \ ns \ ,$$

$$k_{21} = 1 \ /ns \ ,$$

$$k_{02} = 0.1/ns \ .$$

Figure 2 plots $\rho_{00}(\infty)$ and $\rho_{22}(\infty)$ versus $\Omega$ for these parameters. $\rho_{11}(\infty)$ follows from equation (8) or (11) and starts at 0 and then rises to just over 0.072 at $\Omega = 10$ GHz.

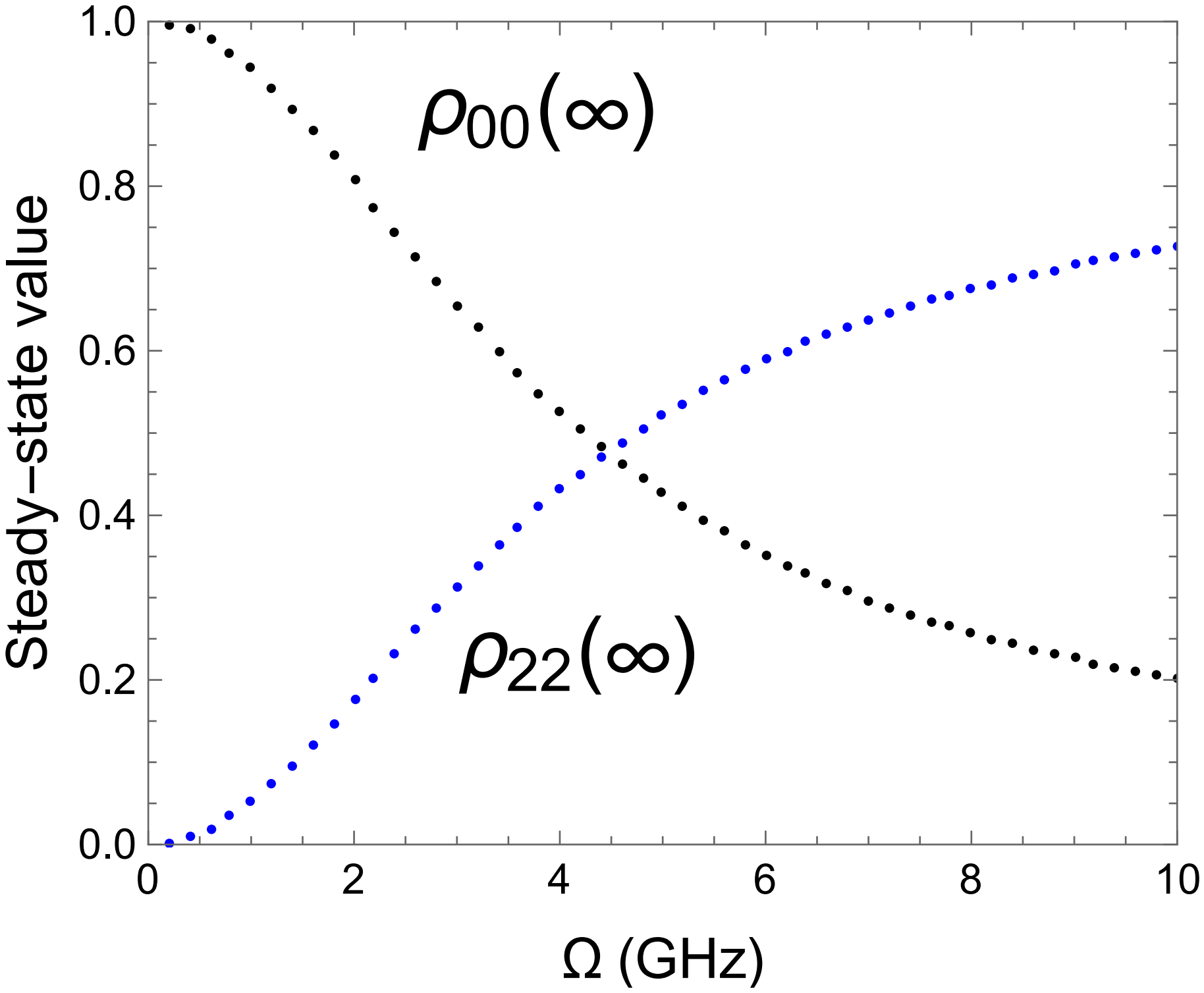

**Figure 2.** The steady-state values for $\rho_{00}(\infty)$ and $\rho_{22}(\infty)$ (blue online) as a function of Ω for the present parameters.

The population is seen to end mostly in the ground state, energy level 0, for weak electric fields. But as Ω grows, the majority of the population ends in energy level 2 for the present parameters. The cross-over occurs here between $\Omega = 4$ and $4.5\times10^9$ radians/s and develops because the transition from energy level 2 to the ground state is rate-limiting. A bottleneck occurs. For completeness, $\rho_B(\infty)$ is plotted in figure 3.

The eigenvalues associated with equations (2) to (5) are considered next. These equations may be viewed as a matrix equation with a solution that is exponential in the time. This leads to

$$\rho_{ij}(t) = \sum_{k=1}^{4} c_{ijk}\, \varphi_k(t) \; , \tag{13}$$

for the three $\rho_{ii}(t)$ and $\rho_B(t)$. In addition,

$$\varphi_k(t) = e^{\gamma_k t} = e^{-t/\tau_k} \; , \tag{14}$$

where the $\gamma_k$ are the eigenvalues and the $\tau_k$ are the decay time constants discussed below. The routine Eigenvalues of Mathematica [15] is used. This section shows the steady-state values of the $\rho_{ii}(t)$ and $\rho_B(t)$ are non-zero. This leads to the expectation that one eigenvalue is zero, otherwise, all the steady-state values are zero. Appendix B shows how this comes about. In addition, the other three eigenvalues are negative or have a negative real part. These conditions are required for each $\rho_{ii}(t)$ and $\rho_B(t)$ to approach its steady-state value as the time *t* grows. Now, since each $\rho_{ij}(t)$ depends on the same eigenvalues, each $\rho_{ij}(t)$ has the same exponential approach to its steady-state value, which is through the smallest nonzero $\gamma_k$.

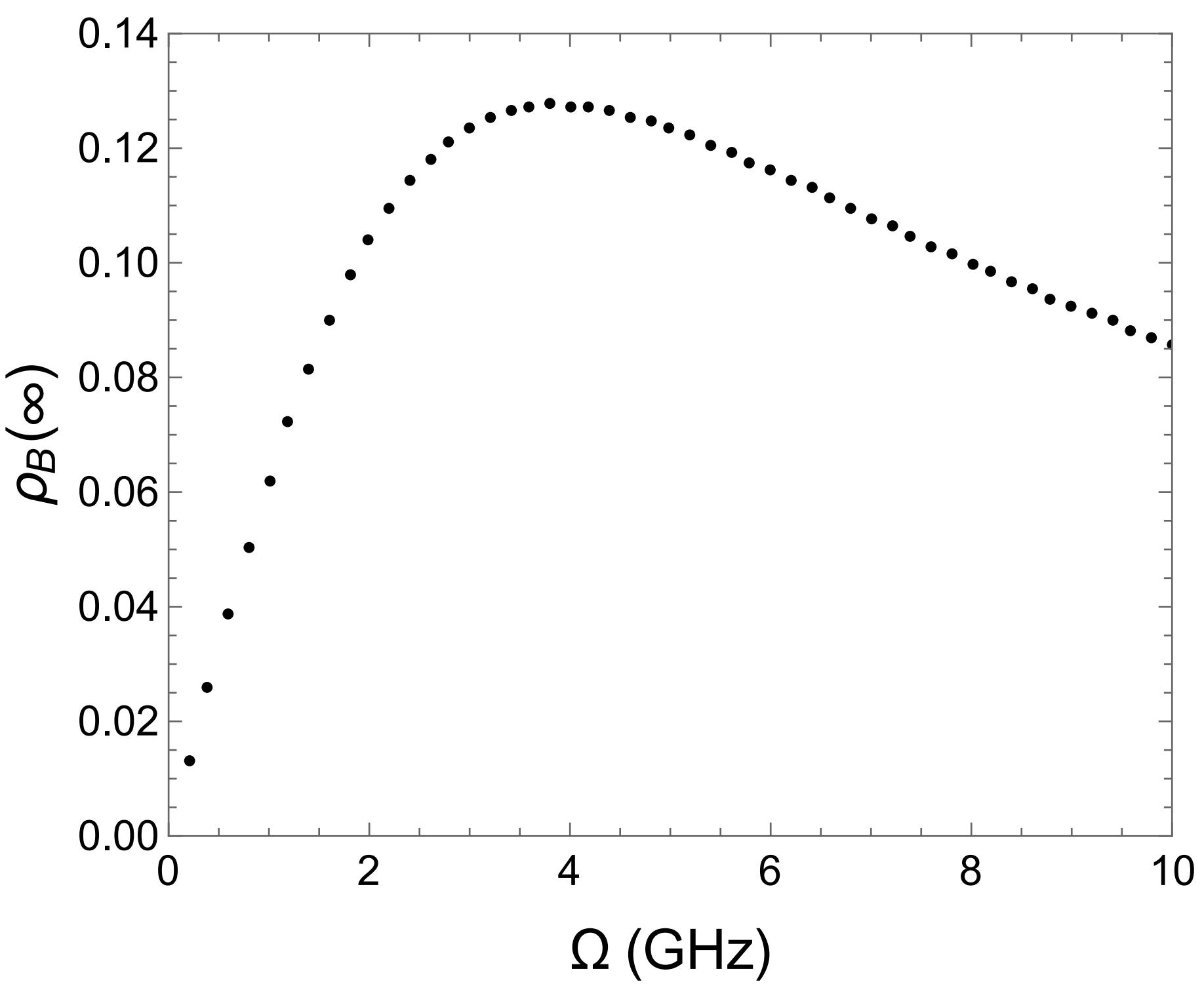

**Figure 3.** The steady-state values for $\rho_B(\infty)$ as a function of $\Omega$ for the present parameters.

The next section probes the time-dependence of the $\rho_{ii}(t)$ and their behavior at both short and long times, makes contact with the eigenvalues, and confirms the above prediction.

## 4. Numerical Results

The initial conditions for the density operator matrix elements are

$$\rho_{00}(t=0) = \rho_{22}(t=0) = \rho_B(t=0) = 0 \,, \tag{15}$$

$$\rho_{11}(t=0) = 1 \,. \tag{16}$$

Thus, the population starts in the excited state of energy level 1 and then leaves this state by stimulated emission to the ground state, energy level 0, or by relaxation to energy levels 0 or 2. This allows the study of the decay of the initial population and the subsequent approach to the steady-state populations. NDSolve of Mathematica [15] is used to provide the numerical solutions of equations (2) to (5). The sum of the occupation probabilities should remain at 1.0 per equation (7) and the numerical solutions are found to satisfy

$$\rho_{00}(t) + \rho_{11}(t) + \rho_{22}(t) = 1.0 \, \pm 0.00001 \,. \tag{17}$$

The first interest is in how the population of $\rho_{11}(t)$ decays with time. Then, the long-time behavior of the density matrix elements is explored. It is found that they approach their steady-state values exponentially, which is in accord with equations (2) to (5). The same decay time constant is extracted from the curves for the four matrix elements using their values at $t = 11$ and $t = 14$ ns. This also agrees with a consideration of the eigenvalues in section 3.

Figure 2 is helpful in selecting the values of Ω to explore. The weak-field regime is represented by $\Omega = 0.1$ GHz, the cross-over is illustrated with $\Omega = 4.5$ GHz, and the stronger-field regime has Ω =10. GHz. The plots of the $\rho_{ii}(t)$ have a solid black line for $\rho_{00}(t)$, a dotted line (blue online) for $\rho_{11}(t)$, and a dashed black line for $\rho_{22}(t)$.

Figure 4 shows how the $\rho_{ii}(t)$ behave with time for the weak-field case with $\Omega = 0.1$ GHz. $\rho_{11}(t)$ decays quickly and this is emphasized in the semi-log plot of figure 5. The time constant is found to be $\tau = 0.084$ ns. Now, in the weak-field case, Appendix B reveals this decay time constant to be

$$ {}^{1}/_{\tau_{eff}} = \left({}^{1}/_{T_1}\right) + k_{21} = 11.83 \text{ ns}^{-1}, \tag{18} $$

so,

$$ \tau_{eff} = 0.0845 \text{ ns}. \tag{19} $$

This reflects the dominance of the two decay modes from the excited state in the weak-field limit and is in excellent agreement with the numerical solutions.

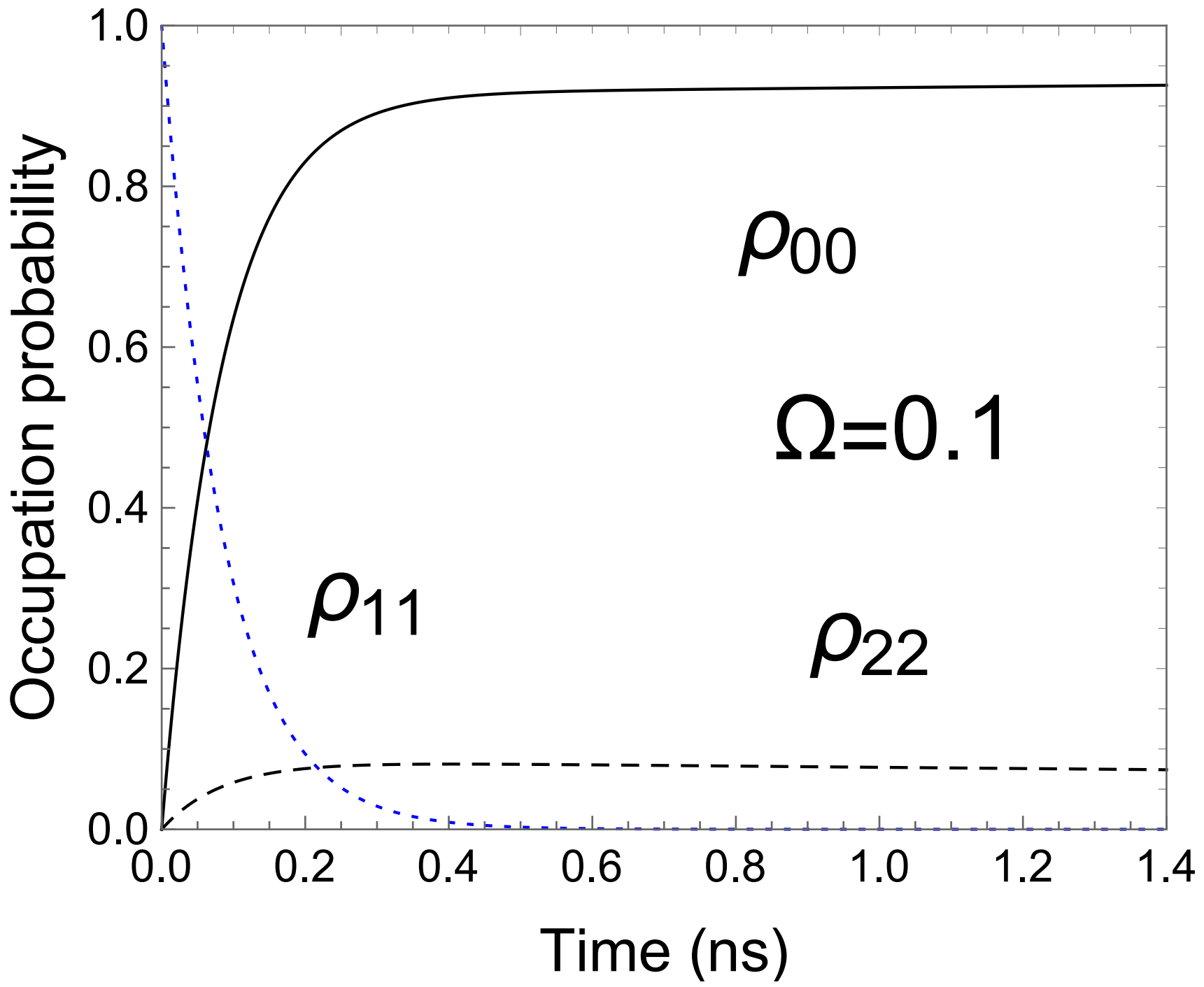

**Figure 4.** The occupation probabilities as functions of time for $\Omega = 0.1$ GHz. The system starts in energy level 1 and resonance is assumed. $\rho_{00}(t)$ is a solid line, $\rho_{11}(t)$ is a dotted line (blue online), and $\rho_{22}(t)$ is a dashed line.

Figure 4 illustrates how $\rho_{00}(t)$ rises with the time, while $\rho_{22}(t)$ slowly increases to a maximum and then starts a slow decay. $\rho_{00}(t)$ and $\rho_{22}(t)$ approach their steady-state values of 0.99939 and 0.0005575, respectively, for the present parameters, while $\rho_B(\infty) = 0.006596$. This long-time behavior is shown in figure 6, where $\rho_{00}(\infty) - \rho_{00}(t)$ , $\rho_{22}(t) - \rho_{22}(\infty)$, and $\rho_B(\infty) - \rho_B(t)$ are plotted versus time. The decay time constant is 9.994 ns, which agrees with the $\tau_3$ based on the eigenvalue $\gamma_3$. A similar plot, which is not shown, for $\rho_{11}(\infty) - \rho_{11}(t)$ yields 10.07 ns. For this case, $\rho_{11}(\infty) = 0.00005575$ and a plot occupies the $10^{-6}$ to $10^{-5}$ decade. A loss of numerical significance probably accounts for the slight difference in the decay time constants. In any case, this time constant is close to the inverse of $k_{02}$, which is $\tau_3 = 10$ ns and is the largest finite decay time constant per Appendix B.

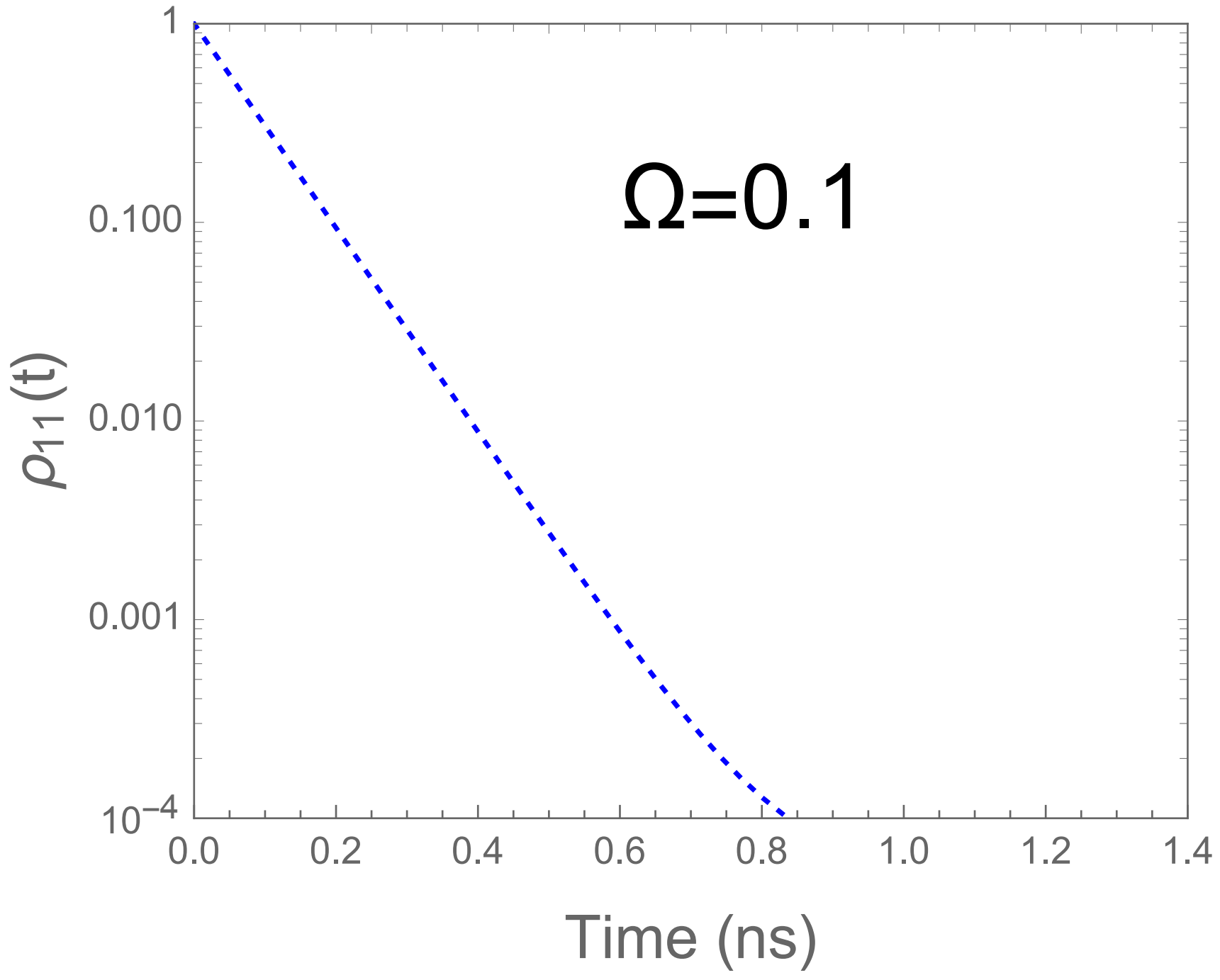

**Figure 5.** The occupation probability $\rho_{11}(t)$ as a function of time for $\Omega = 0.1$ GHz. The system starts in energy level 1 and resonance is assumed. $\rho_{11}(t)$ is a dotted line (blue online). The toe near 0.8 ns is due to $\rho_{11}(t)$ approaching its steady-state value.

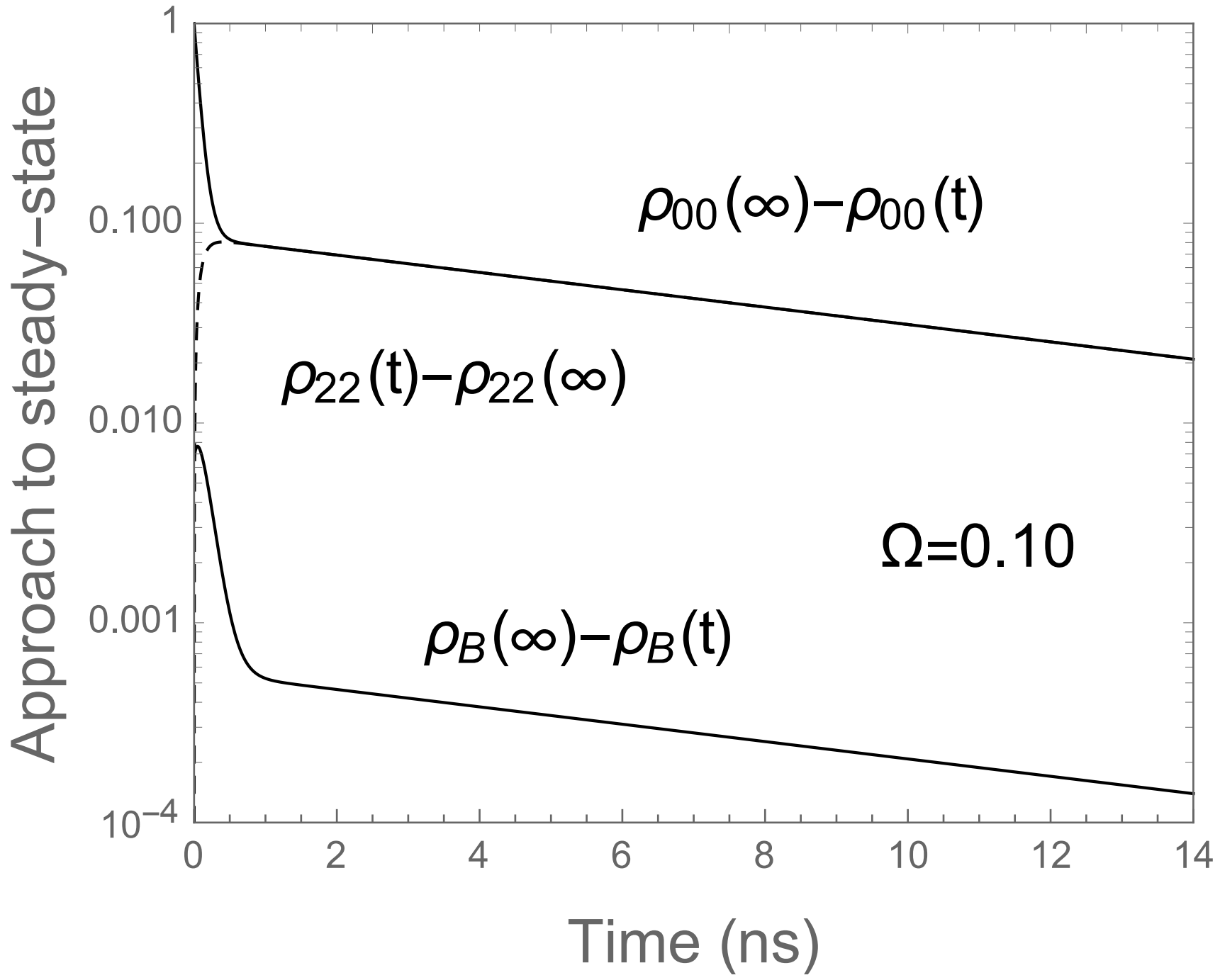

**Figure 6.** The occupation probabilities minus their steady-state values as functions of time for $\Omega = 0.1$ GHz. The system starts in energy level 1 and resonance is assumed. $\rho_{00}(\infty) - \rho_{00}(t)$ and $\rho_B(\infty) - \rho_B(t)$ are solid lines, and $\rho_{22}(t) - \rho_{22}(\infty)$ is a dashed line that merges with $\rho_{00}(\infty) - \rho_{00}(t)$. The time goes from 0 to 14 ns.

The next case has $\Omega = 4.5$ GHz and is roughly where $\rho_{22}(\infty)$ starts to exceed $\rho_{00}(\infty)$ in figure 2. The steady-state values for this case are

$$\rho_{00}(\infty) = 0.4725\ ,$$
$$\rho_{11}(\infty) = 0.04796\ , \qquad (20)$$
$$\rho_{22}(\infty) = 0.4796\ \ ,$$
$$\rho_B(\infty) = 0.1261\ .$$

Figures 7 and 8 display the $\rho_{ii}(t)$ for $t = 0$ to 1.4 and 14 ns, respectively. In both figures, $\rho_{11}(t)$ drops quickly to a minimum and then rises slightly before decreasing towards its steady-state value. In contrast to the $\Omega = 0.1$ GHz case in figure 5, the short-time $\rho_{11}(t)$ decays exponentially over less than 2 decades with a time constant of ~0.07 ns. Figure 8 shows how $\rho_{22}(t)$ approaches $\rho_{00}(t)$. The latter rises rapidly to a peak and then decays to $\rho_{00}(\infty)$, while the former steadily increases to $\rho_{22}(\infty)$. Figure 9 shows how all 4 density operator matrix elements approach their steady-state values with a common decay time constant of 5.117 ns, which differs from that of the first case for $\Omega = 0.1$ GHz. Thus, the decay time constant depends on $\Omega$, but continues to agree with the eigenvalue-based $\tau_3$ defined in equation (14).

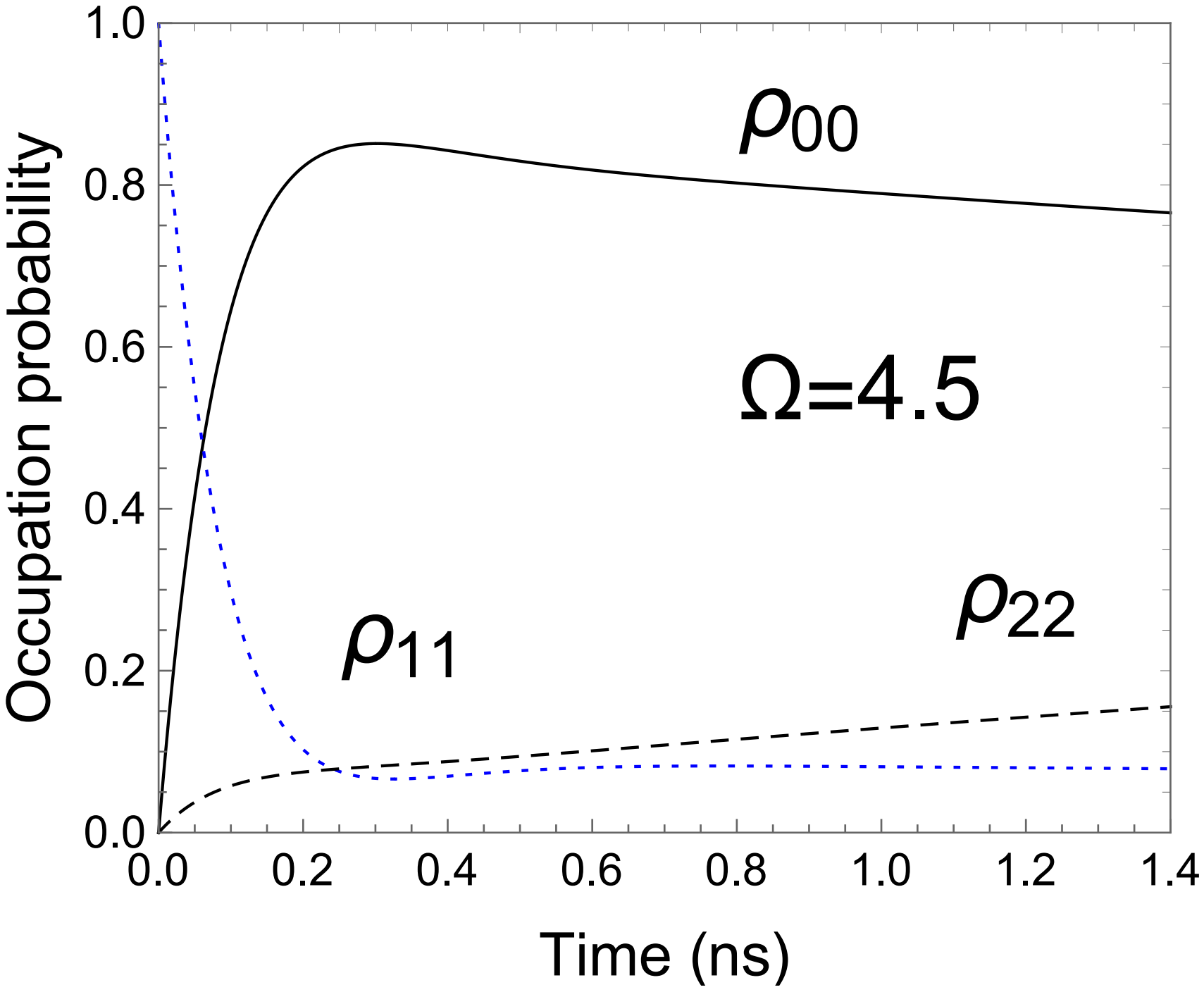

**Figure 7.** The occupation probabilities as functions of time for $\Omega = 4.5$ GHz. The system starts in energy level 1 and resonance is assumed. $\rho_{00}(t)$ is a solid line, $\rho_{11}(t)$ is a dotted line (blue online), and $\rho_{22}(t)$ is a dashed line.

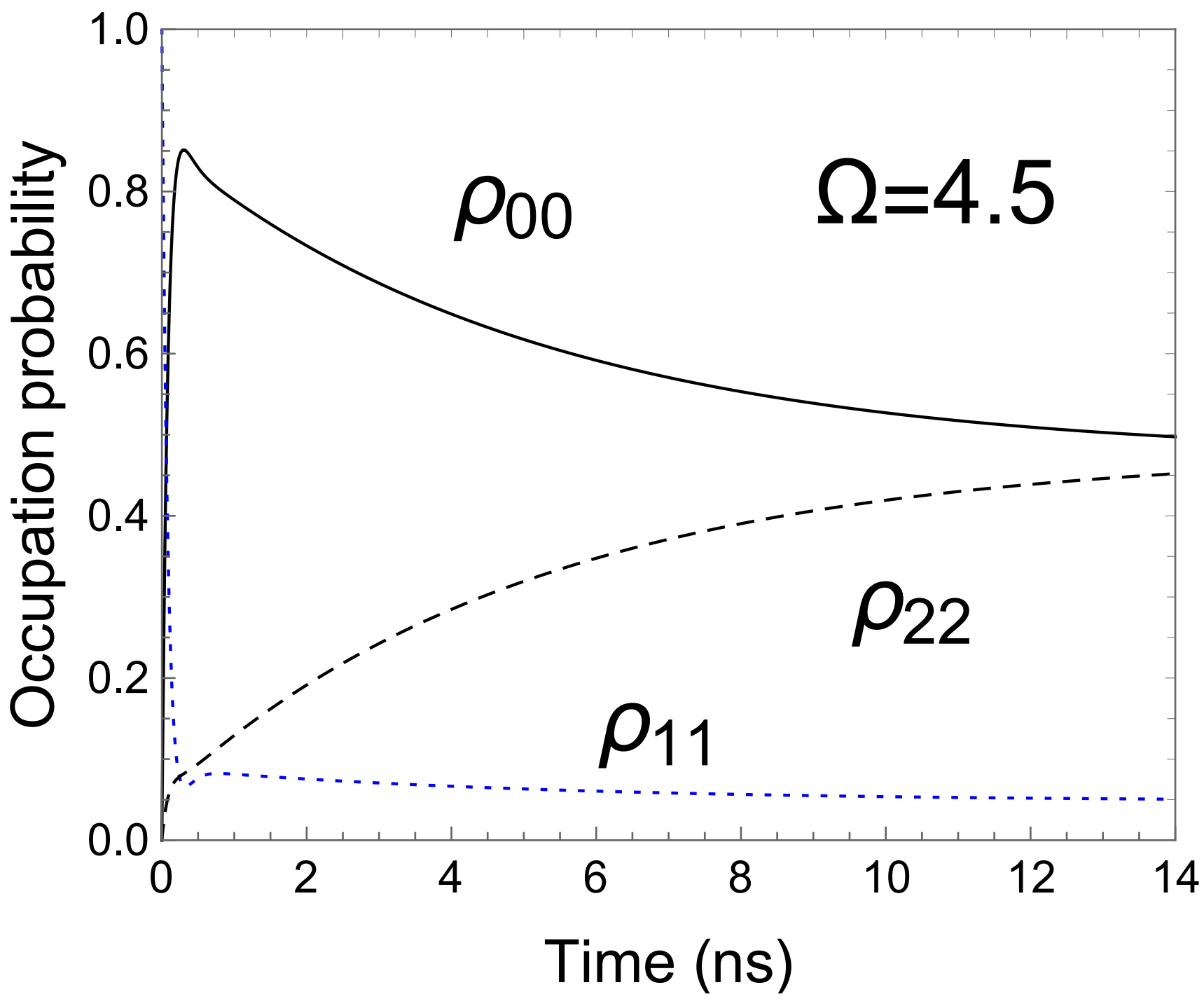

**Figure 8.** The occupation probabilities as functions of time for $\Omega = 4.5$ GHz. The system starts in energy level 1 and resonance is assumed. $\rho_{00}(t)$ is a solid line, $\rho_{11}(t)$ is a dotted line (blue online), and $\rho_{22}(t)$ is a dashed line. The time goes from 0 to 14 ns.

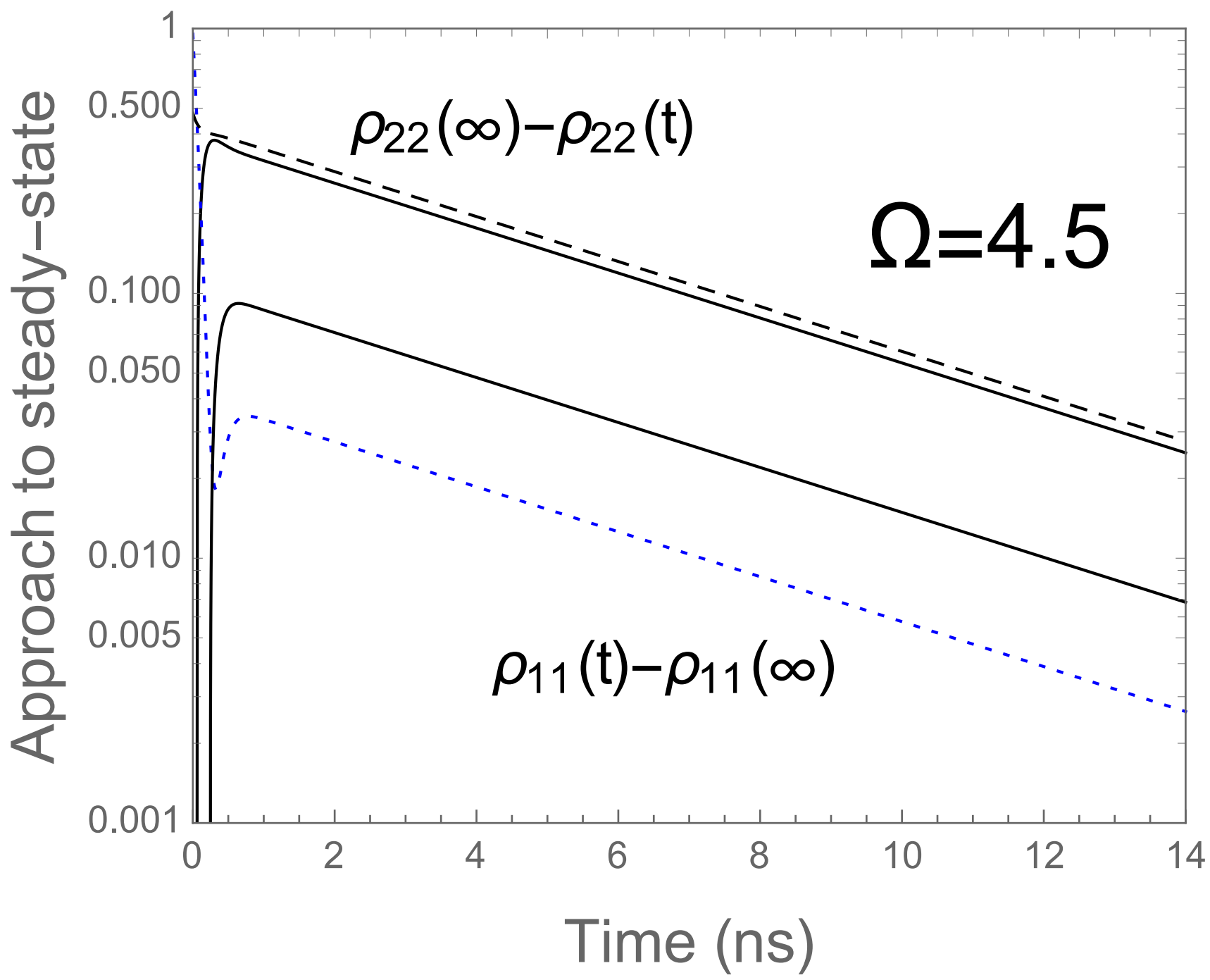

**Figure 9.** The occupation probabilities minus their steady-state values as functions of time for $\Omega = 4.5$ GHz. The system starts in energy level 1 and resonance is assumed. From the top: $\rho_{22}(\infty) - \rho_{22}(t)$ is a dashed line, $\rho_{00}(t) - \rho_{00}(\infty)$ is a solid line, $\rho_{\mathrm{B}}(t) - \rho_B(\infty)$ is a solid line, and $\rho_{11}(t) - \rho_{11}(\infty)$ is a dotted line (blue online).

The final case is for a stronger-field with $\Omega = 10$ GHz and

$$\rho_{00}(\infty) = 0.2025\,,$$

$$\rho_{11}(\infty) = 0.07250\,,$$

$$\rho_{22}(\infty) = 0.7250\,, \tag{21}$$

$$\rho_{B}(\infty) = 0.08577\,.$$

Figures 10 and 11 have the $\rho_{ii}(t)$ for $t = 0$ to 1.4 and 14 ns, respectively. The behaviors seen with the previous case occur here also. The crossover between $\rho_{22}(t)$ and $\rho_{00}(t)$ now happens before 3 ns. Figure 12 has the approach to steady-state of the 4 density operator matrix elements. All 4 plots yield a decay time constant of 2.674 ns, which is $\tau_3$ for this case.

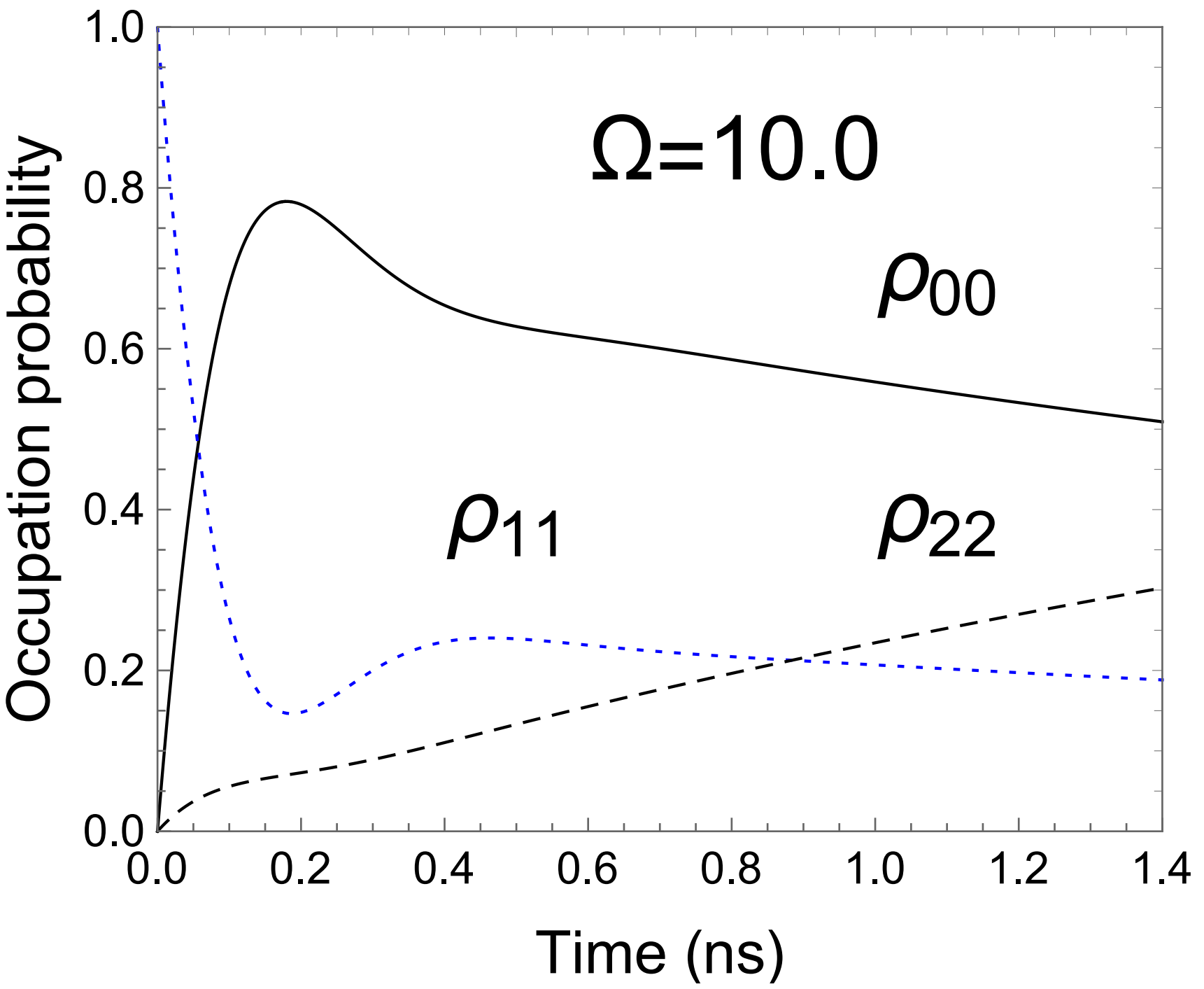

**Figure 10.** The occupation probabilities as functions of time for $\Omega = 10.0$ GHz. The system starts in energy level 1 and resonance is assumed. $\rho_{00}(t)$ is a solid line, $\rho_{11}(t)$ is a dotted line (blue online), and $\rho_{22}(t)$ is a dashed line.

Further results are woven into the concluding section.

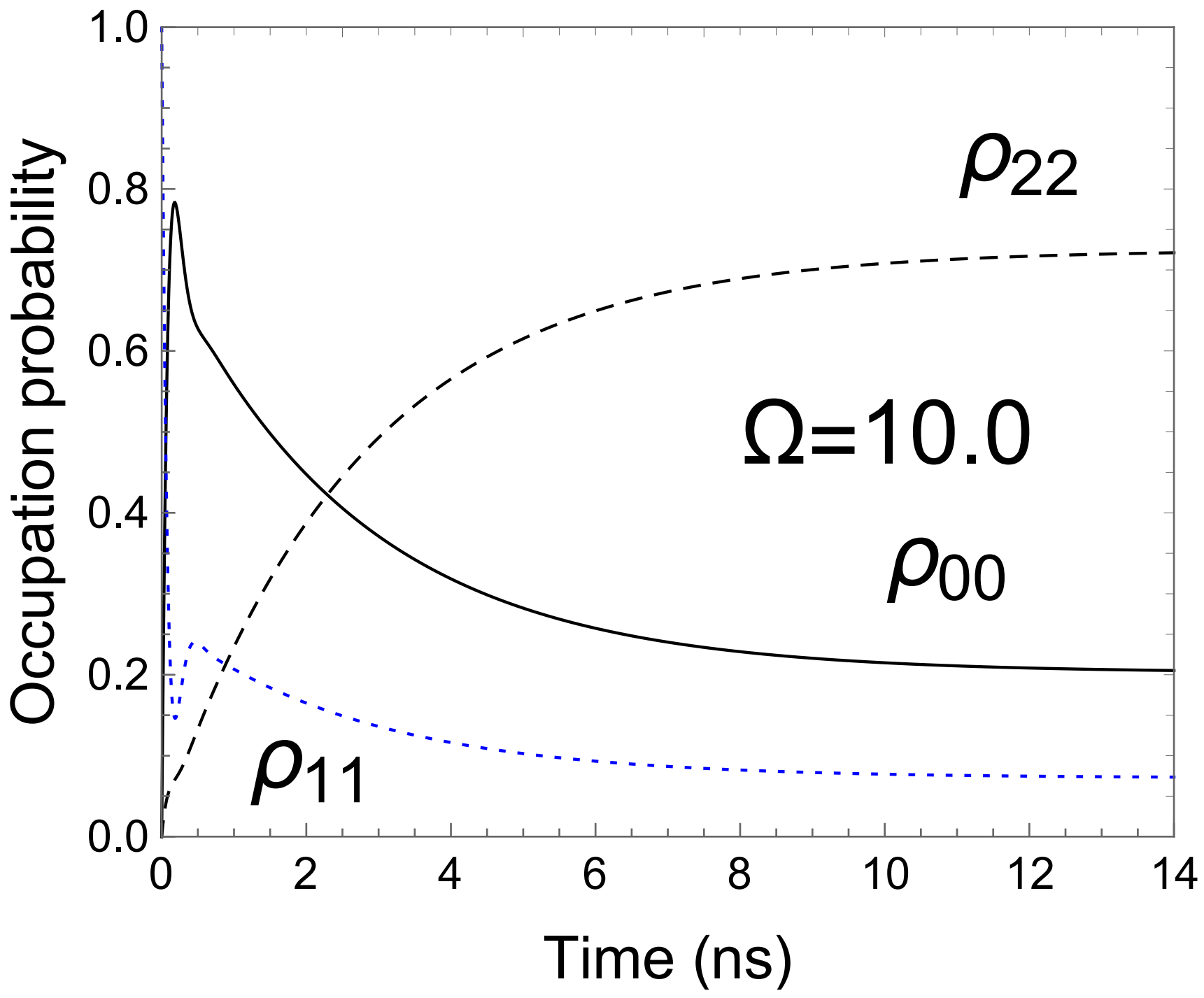

**Figure 11.** The occupation probabilities as functions of time for $\Omega = 10.0$ GHz. The system starts in energy level 1 and resonance is assumed. $\rho_{00}(t)$ is a solid line, $\rho_{11}(t)$ is a dotted line (blue online), and $\rho_{22}(t)$ is a dashed line. The time goes from 0 to 14 ns.

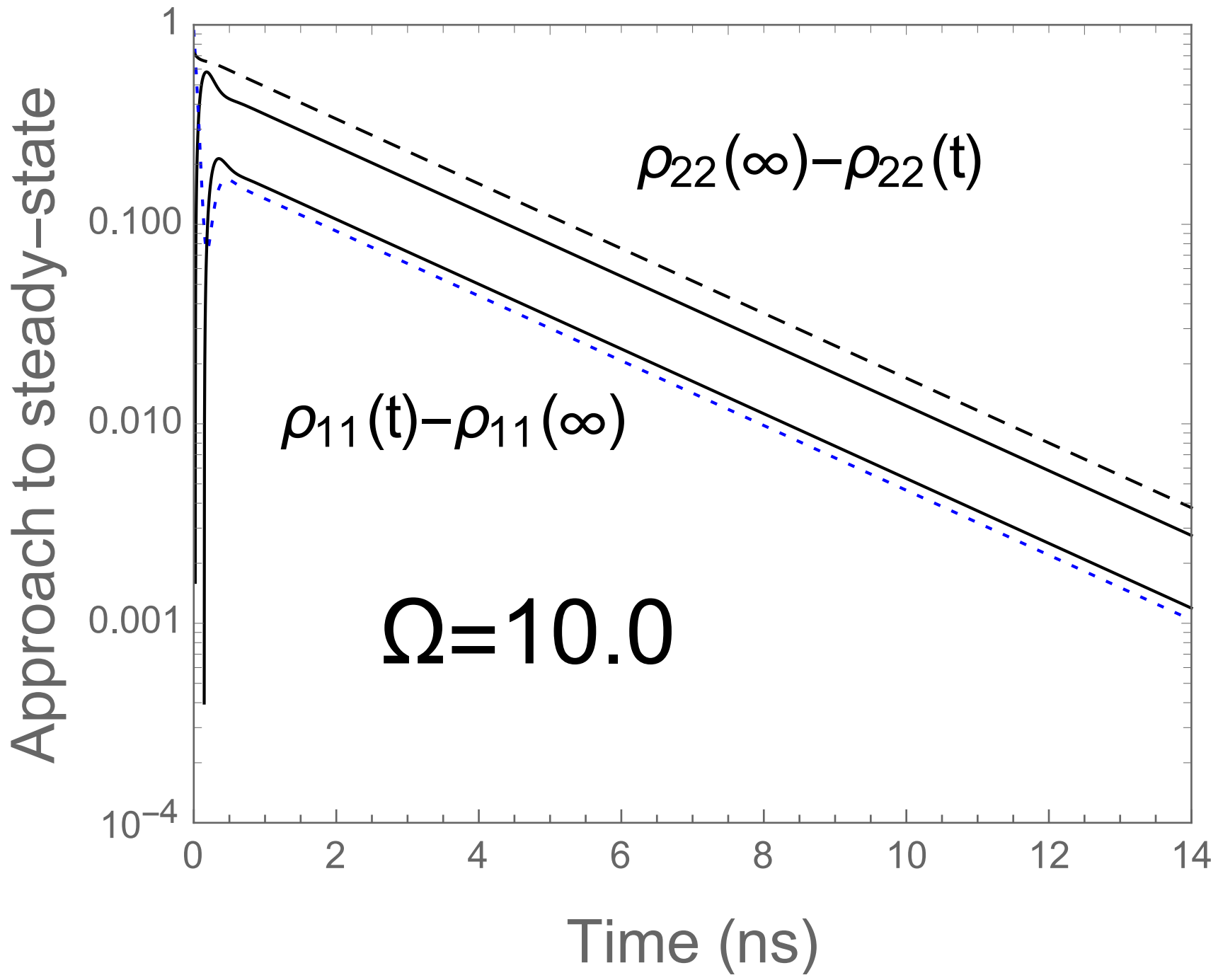

**Figure 12.** The occupation probabilities minus their steady-state values as functions of time for $\Omega = 10.0$ GHz. The system starts in energy level 1 and resonance is assumed. From the top: $\rho_{22}(\infty) - \rho_{22}(t)$ is a dashed line, $\rho_{00}(t) - \rho_{00}(\infty)$ is a solid line, $\rho_{\mathrm{B}}(t) - \rho_{B}(\infty)$ is a solid line, $and$ $\rho_{11}(t) - \rho_{11}(\infty)$ is a dotted line (blue online).

## 5. Conclusions

The previous section explored how the population in the excited state decays and how the $\rho_{ii}(t)$ approach their steady-state values when one leg of the three-level lambda system is driven by a laser at all times. The use of the Interaction Picture, the Rotating Wave Approximation, and resonance allows the eigenvalues to be determined and used to interpret the numerical results. This makes this model system useful for exposing the details of how this system goes to its steady state. In some limits the eigenvalues may be related to the relaxation parameters and the electric field strength.

All 4 density operator matrix elements go to their steady-state values with the same decay time constant that depends on the strength of the laser's electric field magnitude Ω. In addition, this decay constant agrees with that derived from the eigenvalue for that value of Ω via $\tau_3 = -1/\gamma_3$. Figure 13 shows this dependence for the 3 cases treated here as well as for other values of Ω. As seen in equation (13), the eigenvalues of the Liouville-von Neumann equations explain this coincidence of decay time constants for a fixed Ω. Only one time-dependent term survives as the time increases. When Ω goes to zero, this decay time constant approaches $1/k_{02}$ , which is 10 ns here, and almost all of the population ends in the ground state, energy level 0. The increase in Ω leads to a smaller decay time constant that decreases proportionally to a constant minus the logarithm of Ω. The latter behavior is seen in figure 13.

Figure 2 shows that the majority of the population ends in energy level 2 once Ω exceeds approximately 4.4 GHz. This at first appears strange because the lowest energy level is energy level 0. The crossover between $\rho_{22}(\infty)$ and $\rho_{00}(\infty)$ shifts to larger Ω when the rate $k_{02}$ is increased. For example, the crossing occurs at Ω (GHz) equals 6.6, 9.6, and beyond 10.0 for $k_{02}$ (1/ns) equal to 0.20, 0.35, and 0.40, respectively. The transition from energy level 2 to energy level 0 eventually becomes the rate-limiting step. Hence, the population in energy level 2 grows with Ω until it dominates. These changes in $k_{02}$ hardly affect the value of Ω where complex eigenvalues appear.

Finally, at short times and weak electric fields, the population of the initial state, energy level 1, decays with a combination of the relaxation terms as indicated by equation (18). The time extent of this initial decay shrinks with an increase in Ω, since the electric field starts to affect the initial decay through stimulated emission.

This study has explored the time dependence of the occupation probabilities of a three-level lambda system with one leg driven by a laser. The behavior is found by a combination of numerical solutions of the Liouville-von Neumann equations and the eigenvalues of this set of equations. The latter are used to explain the trends found by the numerical solutions. These trends include the decay from the initial state at short times and the approach to the steady-state at long times. In addition, the presence of an eigenvalue equal to zero is shown to occur for all values of Ω and is linked to the non-zero steady-state values of the occupation probabilities. All in all, this model three-level system provides enlightenment for those who probe its behavior and should help students studying time-dependent systems.

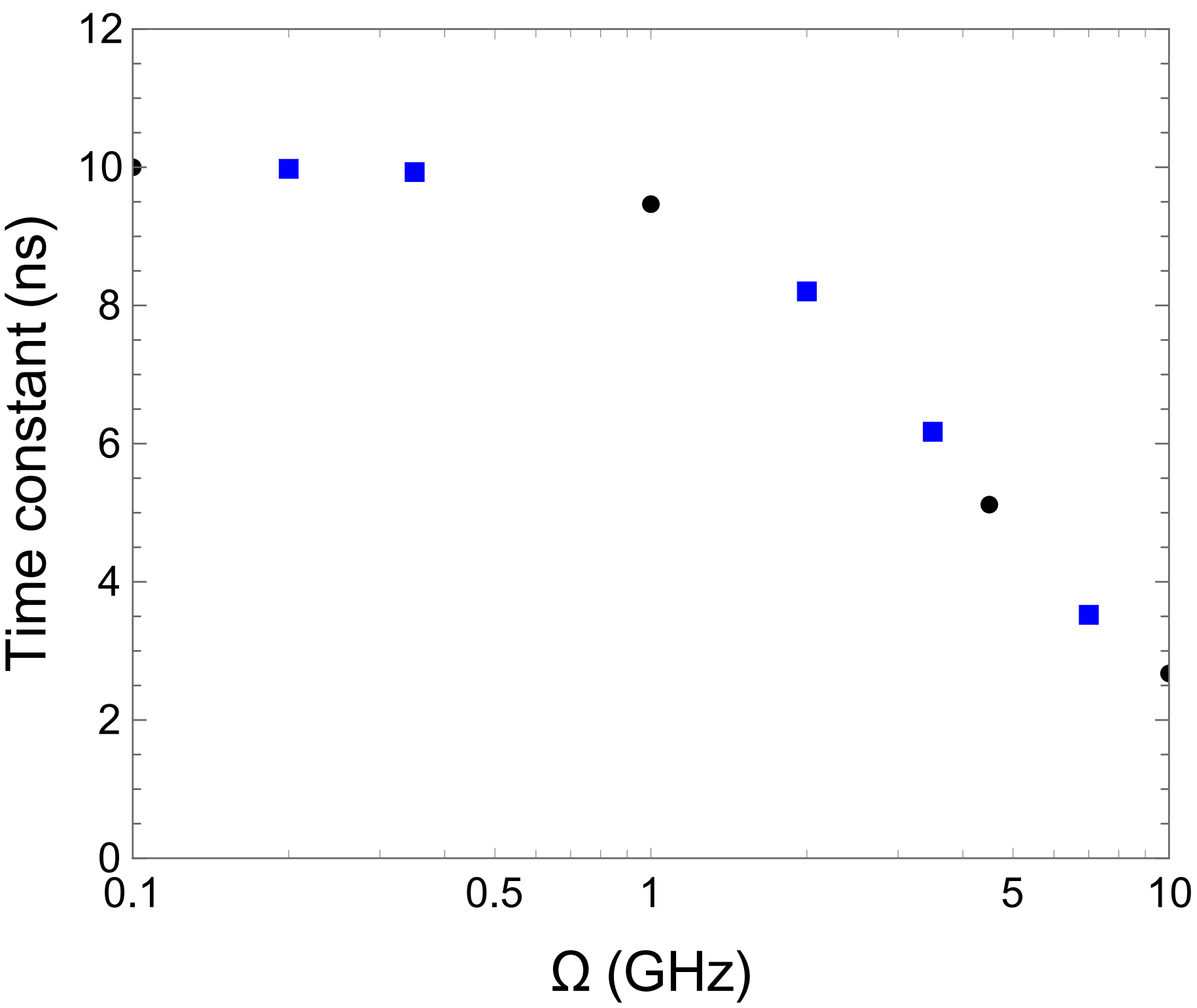

**Figure 13**. The decay time constant as a function of Ω. The solid dots are from the numerical solutions and the solid squares (blue online) are from the calculations of the eigenvalues. The horizontal axis is logarithmic.

## Appendix A

The Liouville-von Neumann equations are developed in the Interaction Picture [2, 3, 12] for numerical work. Briefly, the Hamiltonian is assumed to be

$$H(t) = H_0 + V_c(t) \;, \tag{A1}$$

where $H_0$ provides the energy levels $E_i = \hbar\omega_i$ and

$$V_c(t) = \mu E cos\omega t. \tag{A2}$$

Here $\omega$ is the angular frequency of the laser, which is compared to

$$\Delta\omega = (E_1 - E_0)/\hbar \;. \tag{A3}$$

Now the Interaction Picture matrix element is

$$\langle i|V|j\rangle = \left\langle i\middle|e^{iH_0 t/\hbar}V_c(t)e^{-iH_0 t/\hbar}\middle|j\right\rangle = e^{i\left(E_i - E_j\right)/\hbar}\mu E cos\omega t \;. \tag{A4}$$

Next,

$$cos\omega t = \frac{1}{2}\left(e^{i\omega t} + e^{-i\omega t}\right) \;, \tag{A5}$$

is paired with equation (A4) and only terms with a difference of angular frequencies are retained. This is the Rotating Wave Approximation [2, 4-6]. Finally, the present calculations assume resonance, so $\omega = \Delta\omega$ and the $\langle i|V|j\rangle$ are independent of time.

With the Interaction Picture and the Rotating Wave Approximation, the Liouville-von Neumann Equations are developed from

$$i\hbar\frac{d\rho}{dt} = [V, \rho] \;, \tag{A6}$$

where $V$ and $\rho$ are the potential and the density operator in the Interaction Picture, respectively. Relaxation terms are later added to the equations for the density operator matrix elements, which start from

$$i\hbar\left\langle j\middle|\frac{d\rho}{dt}\middle|l\right\rangle = i\hbar\langle j|\dot{\rho}|l\rangle = \sum_{k=0}^{2}\langle j|V|k\rangle\langle k|\rho|l\rangle - \sum_{k=0}^{2}\langle j|\rho|k\rangle\langle k|V|l\rangle \;. \tag{A7}$$

It is assumed that

$$\langle j|V|j\rangle = 0 \;, \tag{A8}$$

and

$$\langle j|V|k\rangle = 0\ , \tag{A9}$$

unless *j* and *k* are 0 or 1. The non-zero values of equation (A9) correspond to the terms involving the laser that causes transitions between the ground state, energy level 0, and the excited state, energy level 1.

With these rules, the time-dependent equations for the density matrix elements are

$$i\hbar\dot{\rho}_{00} = i\hbar\langle 0|\dot{\rho}|0\rangle = \langle 0|V|1\rangle\rho_{10} \ - \ \rho_{01}\langle 1|V|0\rangle\ , \tag{A10}$$

$$i\hbar\dot{\rho}_{01} = i\hbar\langle 0|\dot{\rho}|1\rangle = \langle 0|V|1\rangle\rho_{11} - \rho_{00}\langle 0|V|1\rangle\ , \tag{A11}$$

$$i\hbar\dot{\rho}_{02} = i\hbar\langle 0|\dot{\rho}|2\rangle = \langle 0|V|1\rangle\rho_{12}\ , \tag{A12}$$

$$i\hbar\dot{\rho}_{10} = i\hbar\langle 1|\dot{\rho}|0\rangle = \langle 1|V|0\rangle\rho_{00} - \rho_{11}\langle 1|V|0\rangle\ , \tag{A13}$$

$$i\hbar\dot{\rho}_{11} = i\hbar\langle 1|\dot{\rho}|1\rangle = \langle 1|V|0\rangle\rho_{01} \ - \ \rho_{10}\langle 0|V|1\rangle\ , \tag{A14}$$

$$i\hbar\dot{\rho}_{12} = i\hbar\langle 1|\dot{\rho}|2\rangle = \langle 1|V|0\rangle\rho_{02}\ , \tag{A15}$$

$$i\hbar\dot{\rho}_{20} = i\hbar\langle 2|\dot{\rho}|0\rangle = -\rho_{21}\langle 1|V|0\rangle\ , \tag{A16}$$

$$i\hbar\dot{\rho}_{21} = i\hbar\langle 2|\dot{\rho}|1\rangle = -\rho_{20}\langle 0|V|1\rangle\ , \tag{A17}$$

$$i\hbar\dot{\rho}_{22} = 0\ . \tag{A18}$$

Equations (A12) and (A15) are coupled as are equations (A16) and (A17). Neither set involves the occupation probabilities. In addition to equation (A18), the sum of equations (A10) and (A14) is zero, so probability is conserved

$$\frac{\partial}{\partial t}\big(\rho_{00}(t) + \rho_{11}(t) + \rho_{22}(t)\big) = 0\ . \tag{A19}$$

The relaxation terms that are added preserve equation (A19).

The four equations (A10), (A11), (A13), and (A14) form a coupled set. The terms with $\rho_{01}$ and $\rho_{10}$ are combined into a term for the imaginary part of their difference, the $\rho_B$ of Section II. No matrix elements of *V* involve the index 2, so the equation for $\dot{\rho}_{22}$ has only relaxation terms, which couple it $\rho_{11}$. In addition, $\rho_{22}$ is coupled to $\rho_{00}$ , to which it relaxes. Finally, all of this leads to equations (2) to (5), when all the relaxation terms are introduced.

## Appendix B

Equation (14), which is rewritten here as equation (B1), turns equations (2) to (5) into an

$$\varphi(t) = e^{\gamma t} = e^{-t/\tau}, \tag{B1}$$

eigenvalue problem for $\gamma$ and suggests two tasks. The first is to see why one of the eigenvalues is zero and the second is to find the eigenvalues in the limit of $\Omega$ approaching zero.

The eigenvalues come from solving

$$det\, M \;=\; 0\,, \tag{B2}$$

with

$$M \;=\; \begin{vmatrix} -\gamma & -\Omega & {}^{1}/_{T_1} & k_{02} \\ {}^{\Omega}/_{2} & -\gamma - \left({}^{1}/_{T_2}\right) & -{}^{\Omega}/_{2} & 0 \\ 0 & \Omega & -\left({}^{1}/_{T_1}\right) - k_{21} - \gamma & 0 \\ 0 & 0 & k_{21} & -k_{02} - \gamma \end{vmatrix}. \tag{B3}$$

Here the order is $\rho_{00}$, $\rho_{B}$, $\rho_{11}$, and $\rho_{22}$. When row 3 is added to row 1,

$$M \;=\; \begin{vmatrix} -\gamma & 0 & -k_{21} - \gamma & k_{02} \\ {}^{\Omega}/_{2} & -\gamma - \left({}^{1}/_{T_2}\right) & -{}^{\Omega}/_{2} & 0 \\ 0 & \Omega & -\left({}^{1}/_{T_1}\right) - k_{21} - \gamma & 0 \\ 0 & 0 & k_{21} & -k_{02} - \gamma \end{vmatrix}. \tag{B4}$$

Now, the determinant is found by expanding with column 1,

$$det\, M \;=\; -\gamma\, detM_1 - \left({}^{\Omega}/_{2}\right) detM_2\,. \tag{B5}$$

The question is whether the second term goes as $\gamma$. Now

$$det\ M_2 = \begin{vmatrix} 0 & -k_{21} - \gamma & k_{02} \\ \Omega & -\left({}^{1}/_{T_1}\right) - k_{21} - \gamma & 0 \\ 0 & k_{21} & -k_{02} - \gamma \end{vmatrix} = -\gamma\Omega(k_{02} + k_{21} + \gamma)\ . \tag{B6}$$

Thus, $det\ M\ =\ 0$, yields one eigenvalue equal to 0, which is called $\gamma_4$ here. The eigenvalues are numbered from the most negative real part to zero.

The second task is to find the limit of the eigenvalues when Ω approaches zero. Equation (B5) leads to

$$det\ M \sim -\ \gamma det M_1\ , \tag{B7}$$

with the determinant in this limit going to

$$det\ M_1 = \begin{vmatrix} -\gamma - \left({}^{1}/_{T_2}\right) & 0 & 0 \\ 0 & -\left({}^{1}/_{T_1}\right) - k_{21} - \gamma & 0 \\ 0 & k_{21} & -k_{02} - \gamma \end{vmatrix}. \tag{B8}$$

Setting this determinant to zero, reveals

$$\gamma_2 = -{}^{1}/_{T_2}\ . \tag{B9}$$

The lower right 2 x 2 in equation (B8) leads to a quadratic equation in $\gamma$, which gives, after some algebra,

$$\gamma_3 = -k_{02}\ , \tag{B10}$$

and

$$\gamma_1 = -{}^{1}/_{T_1} - k_{21}\ . \tag{B11}$$

The present parameters are given in section 3 and these lead to

$$\gamma_1 = -\left(\frac{1}{0.923333} + 1\right) = -11.834\ ,\ \gamma_2 = -{}^{1}/_{0.132} = -7.57576\ ,\ \gamma_3 = -0.1\ \ . \tag{B12}$$

These are in 1/ns and they become decay time constants through $\tau = -{}^{1}/_{\gamma}$. Thus, in the limit of $\Omega = 0$,

$$\tau_1 = 0.0845285 \text{ ns}, \quad \tau_2 = 0.132 \text{ ns}, \quad \tau_3 = 10 \text{ ns}. \tag{B13}$$

The eigenvalues in the limit of Ω going to zero are given in table 1.

Table 1
Eigenvalues of the Liouville-von Neumann Equations in the weak-field limit

| Ω (GHz) | $\gamma_1$ (1/ns) | $\gamma_2$ (1/ns) | $\gamma_3$ (1/ns) | $\gamma_4$ (1/ns) |
|---|---|---|---|---|
| 1.0 | -11.5922 | -7.80826 | -0.105644 | 0.0 |
| 0.1 | -11,8281 | -7.57795 | -0.100057 | 0.0 |
| 0.01 | -11.8303 | -7.57578 | 0.100001 | 0.0 |
| 0.001 | -11.8303 | -7.57576 | -0.1 | 0.0 |
| 0.0001 | -11.8303 | -7.57576 | -0.1 | 0.0 |

The eigenvalues sum to the trace of the matrix *M*, and for the present parameters

$$\text{Trace}(M) = -\left({}^{1}/_{T_2} + {}^{1}/_{T_1} + k_{21} + k_{02}\right) = -19.50608 \;. \tag{B14}$$

This relation serves as a check on the eigenvalues found numerically and is satisfied for all the cases reported here.

Eigenvalues $\gamma_1$ and $\gamma_2$ start out real for small values of Ω and then go complex around $\Omega = 2.185$ GHz as shown in figure 14. Here both eigenvalues are plotted until they become a complex conjugate pair, and then the real part is plotted. The absolute value of the imaginary part is displayed in figure 15. Its rise with Ω becomes approximately linear after about $\Omega = 4.5$ GHz. Figure 16 depicts the rise of the third eigenvalue, $\gamma_3$, with Ω. The negative of the eigenvalue is plotted and the rise approaches $\Omega^{0.81}$ and then starts to level off for larger Ω.

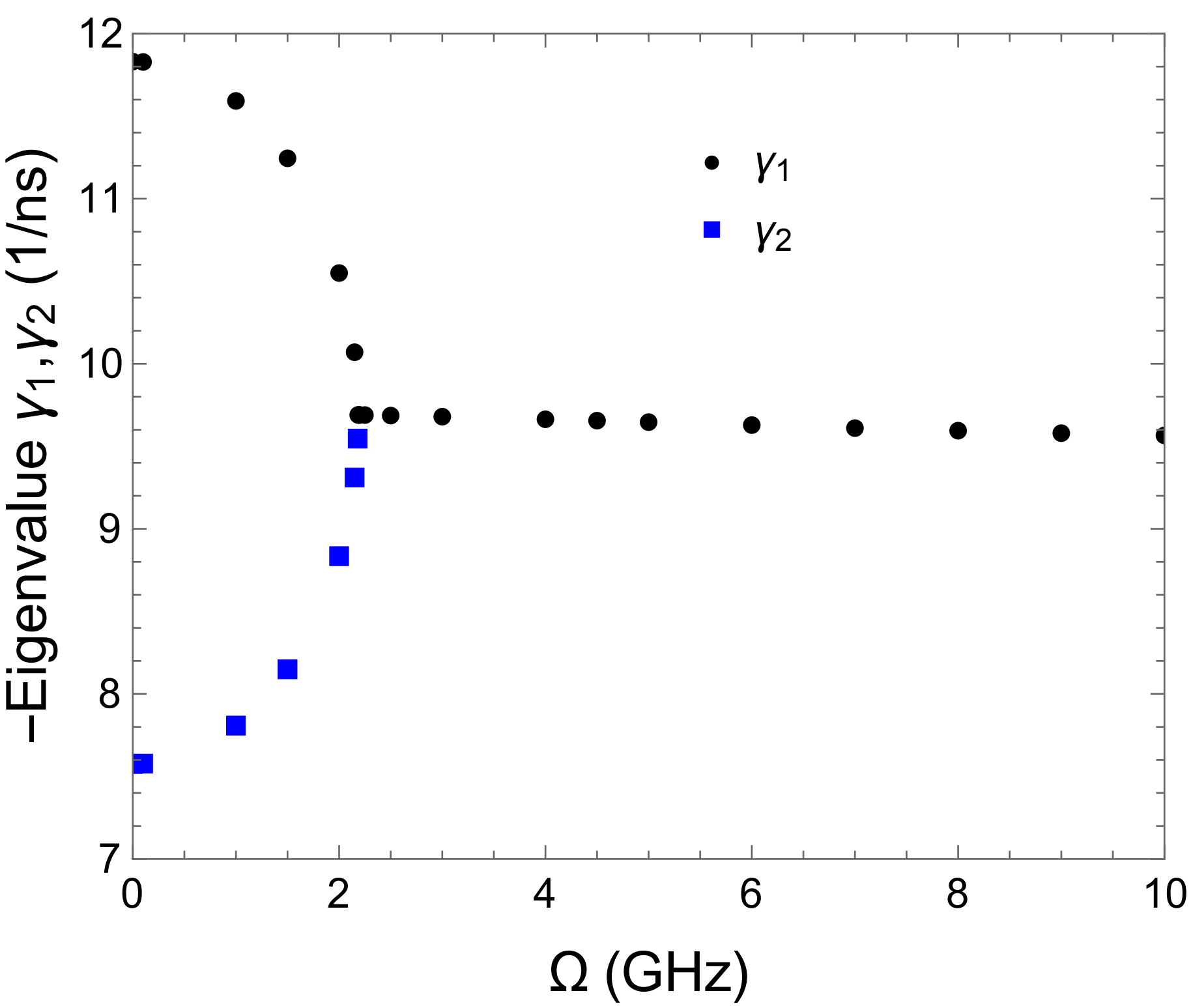

**Figure 14.** The negatives of the real parts of the eigenvalues $\gamma_1$ and $\gamma_2$ are plotted as functions of Ω. Solid dots for $\gamma_1$ and solid squares (blue online) for $\gamma_2$. After both eigenvalues become complex, only their real part is shown.

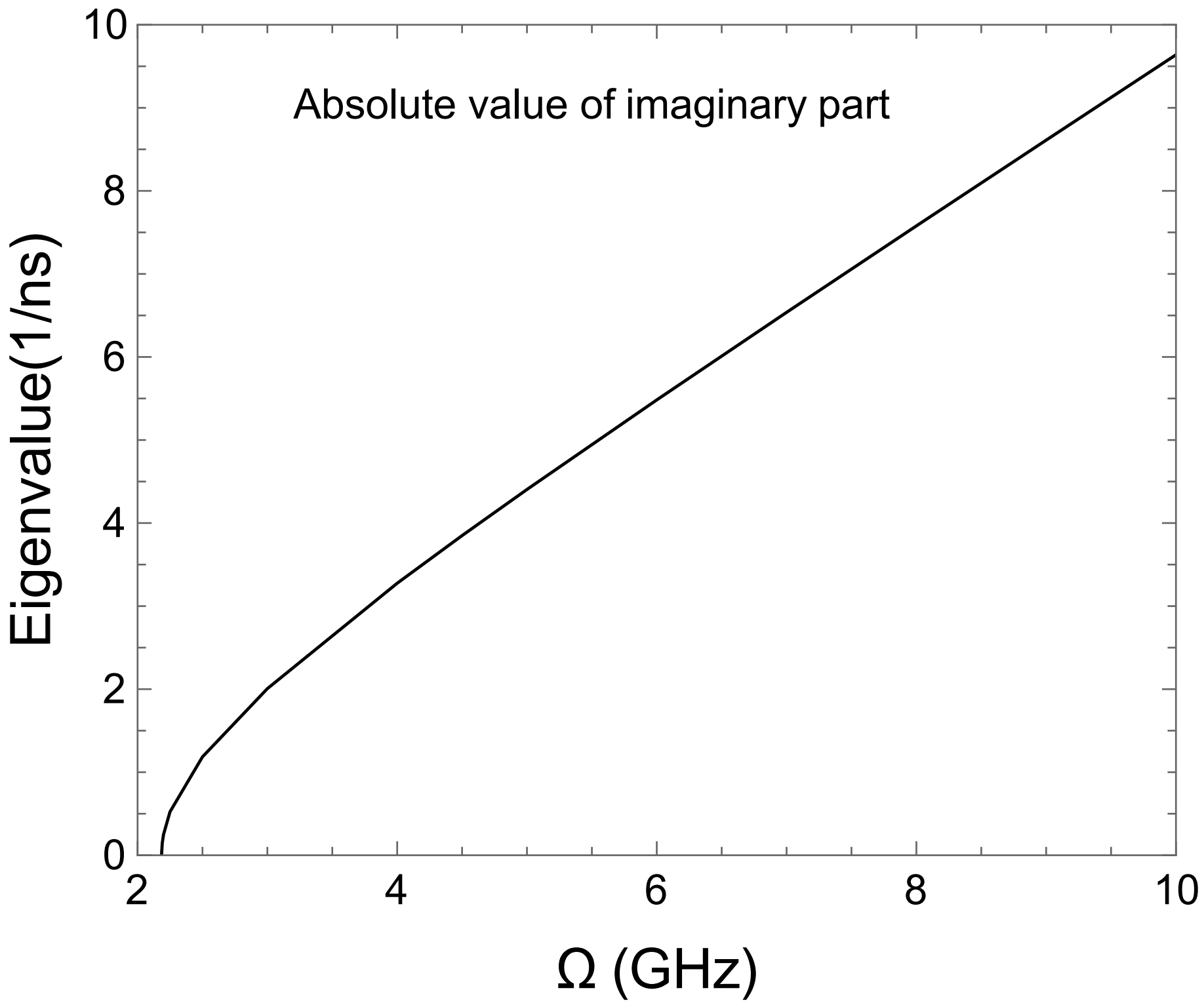

**Figure 15.** The imaginary part of the eigenvalues $\gamma_1$ and $\gamma_2$ is plotted as an absolute value versus Ω. Please note the abscissa starts at 2 GHz.

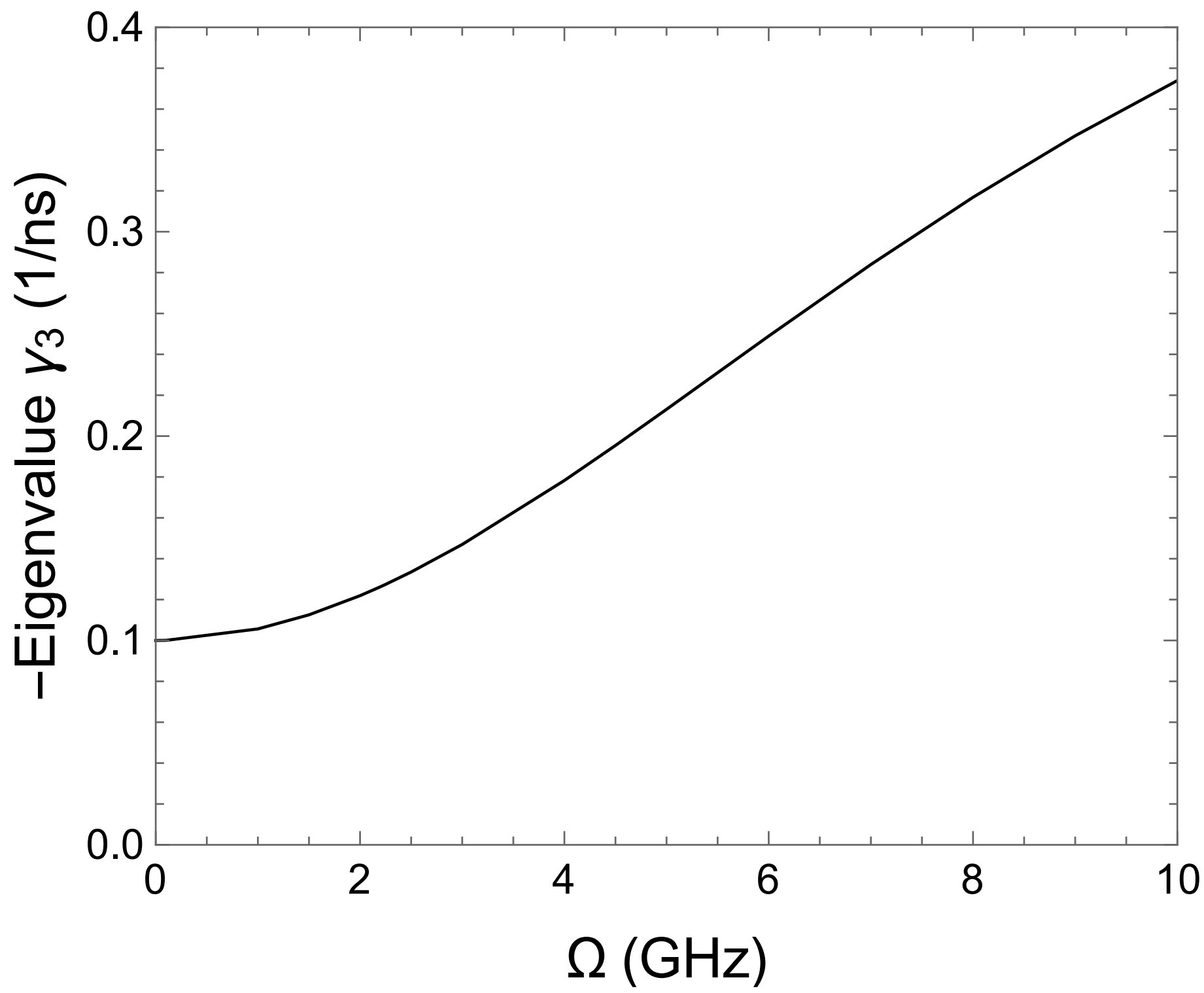

**Figure 16**. The rise of $-\gamma_3$ with $\Omega$. The growth is a power law with an exponent of 0.81 for the larger $\Omega$-values plotted here.

## Acknowledgments

I thank James K. Freericks for asking a question about fast versus slow decay that eventually led to this work.

## Data Availability

The data that support the findings of this study are available within the article.

### ORCID ID

James P Lavine iD https://orcid.org/0000-0002-7337-7082

## References

[1] Van Vliet C M 2008 *Equilibrium and Non-Equilibrium Statistical Mechanics* (Singapore: World Scientific Publishing)

[2] Bransden B H and Joachain C J 2000 *Quantum Mechanics* 2nd edn ( Harlow, England: Prentice Hall)

[3] Sakurai J J and Napolitano J 2021 *Modern Quantum Mechanics* 3rd edn (Cambridge: Cambridge University Press)

[4] Berman P R and Malinovsky V S 2011 *Principles of Laser Spectroscopy and Quantum Optics* (Princeton: Princeton University Press)

[5] Shore B W 1990 *The Theory of Coherent Atomic Excitation* vol 2 (New York: John Wiley and Sons)

[6] Scully M O and Zubairy M S 1997 *Quantum Optics* (Cambridge: Cambridge University Press)

[7] Sanchez B N and Brandes T 2004 Matrix perturbation theory for driven three-level systems with damping *Ann. Phys. (Leipzig)* **13** 569-594

[8] Manka A S, Doss H M, Narducci L M, Ru P, and Oppo G-L 1991 Spontaneous emission and absorption properties of a driven three-level system. II. The Λ and cascade models Phys. Rev A**43** 3748-3763

[9] Blaauboer M 1997 Steady-state behavior in atomic three-level Λ and ladder systems with incoherent population pumping Phys. Rev. A**55** 2459-2462

[10] Rose H, Popolitova D V, Tikhonova O V, Meier T, and Sharapova P R 2021 Dark-state and loss-induced phenomena in the quantum-optical regime of Λ-type three-level systems Phys. Rev. A**103** 013702

[11] Sen S, Dey T K, Nath M R, and Gangopadhyay G 2014 Comparison of electromagnetically induced transparency in lambda, cascade and vee three-level systems J. Mod. Opt. **62** 166-174

[12] Lavine J P 2019 *Time-Dependent Quantum Mechanics of Two-Level Systems* ch 5 (Hackensack, NJ: World Scientific Publishing)

[13] Blum K 2012 *Density Matrix Theory and Applications* 3rd edn (New York: Springer-Verlag)

[14] Basché T, Kummer S, and Bräuchle C 1997 Excitation and emission spectroscopy and quantum optical measurements in *Single-Molecule Optical Detection, Imaging and Spectroscopy* pp 31-67 edited by Basché T, Moerner W E, Orrit M, and Wild U P (Weinheim: VCH Verlagsgesellschaft)

[15] Mathematica = Wolfram Research, Inc., Mathematica, Version 12.0, Champaign, IL (2019).